\documentclass[letter]{aa}
\usepackage{txfonts}
\usepackage{graphicx}
\usepackage{natbib}
\usepackage{amssymb}
\usepackage{lscape}
\bibpunct{(}{)}{;}{a}{}{,}

\def\chandra{{\it Chandra~}}

\def\swift{{\it Swift~}}

\def\m31{{M~31}}
\def\msun{{$M_{\sun}$}}

\def\md{HPH2013~}
\def\mdk{HPH2013}

\newcommand{\nh}{\hbox{$N_{\rm H}$}~}
\newcommand{\hcm}[1]{$\times 10^{#1}$ cm$^{-2}$}

\newcommand{\tpower}[1]{$\times 10^{#1}$}
\newcommand{\power}[1]{$10^{#1}$}

\newcommand{\ton}{$t_{\mbox{\small{on}}}$~}

\newcommand{\vexp}{$v_{\mbox{\small{exp}}}$~}
\newcommand{\ttwo}{$t_{2,R}$~}

\newcommand{\tonk}{$t_{\mbox{\small{on}}}$}
\newcommand{\toffk}{$t_{\mbox{\small{off}}}$}

\def\nova{{M31N~2008-12a~}}
\def\novak{{M31N~2008-12a}}

\def\optk{{DWB2014}}

\begin{document}

\title{A remarkable recurrent nova in \m31 - The X-ray observations}

\author{M.~Henze\inst{1}
	\and J.-U.~Ness\inst{1}
  \and M.~J.~Darnley\inst{2}
	\and M.~F.~Bode\inst{2}
  \and S.~C.~Williams\inst{2}
  \and A.~W.~Shafter\inst{3}
  \and M.~Kato\inst{4}
  \and I.~Hachisu\inst{5}
}

\institute{European Space Astronomy Centre, P.O. Box 78, 28692 Villanueva de la Ca\~{n}ada, Madrid, Spain\\
	email: \texttt{mhenze@sciops.esa.int}
	\and Astrophysics Research Institute, Liverpool John Moores University, IC2 Liverpool Science Park, Liverpool, L3 5RF, UK
  \and Department of Astronomy, San Diego State University, San Diego, CA 92182, USA
  \and Department of Astronomy, Keio University, Hiyoshi, Yokohama 223-8521, Japan
  \and Department of Earth Science and Astronomy, College of Arts and Sciences, The University of Tokyo, Komaba, Meguro-ku, Tokyo 153-8902, Japan
}

\date{Received 13 January 2014 / Accepted 7 February 2014}

\abstract
{Another outburst of the recurrent \m31 nova \nova was announced in late November 2013. Optical data suggest an unprecedentedly short recurrence time of approximately one year.}
{In this Letter we address the X-ray properties of \novak.}
{We requested \swift monitoring observations shortly after the optical discovery. We estimated source count rates and extracted X-ray spectra from the resulting data. The corresponding ultraviolet (UV) data was also analysed.}
{\nova was clearly detected as a bright supersoft X-ray source (SSS) only six days after the well-constrained optical discovery. It displayed a short SSS phase of two weeks duration and an exceptionally hot X-ray spectrum with an effective blackbody temperature of $\sim97$~eV. During the SSS phase the X-ray light curve displayed significant variability that might have been accompanied by spectral variations. The very early X-ray variability was found to be anti-correlated with simultaneous variations in the UV flux.}
{The X-ray properties of \nova coherently point towards a high-mass white dwarf in the nova system. This object might be a promising Type Ia supernova progenitor. We re-discovered additional X-ray detections of \nova that are consistent with our data and increase the number of known nova outbursts to seven. This nova is an exceptional object that merits further attention in the future.}


\titlerunning{Remarkable \m31 RN in X-rays}

\maketitle

%
%
\section{Introduction}
\label{sec:intro}
%
Recurrent Novae (RNe) are those for which more than one nova outburst has been observed. As in the case for Classical Novae, the outburst is caused by a thermonuclear runaway in an accreted envelope of matter on top of a white dwarf (WD) in a close binary system \citep[see][for recent reviews]{2008clno.book.....B}. A proportion of the accreted matter is ejected during the outburst and forms a drastically enlarged pseudo-photosphere, thereby leading to the optical nova. The matter remaining on the WD hosts stable hydrogen burning soon after the outburst. This process powers a supersoft X-ray source \citep[SSS; photon energies below 1~keV, see][]{1998A&A...332..199P} that can be observed once the ejected matter becomes optically thin to soft X-rays.

While the optical characteristics of a nova depend on the properties of the ejected envelope, only X-ray measurements allow us to observe the WD directly and study its physics. Novae with a short SSS phase are expected to host a massive WD \citep[e.g.][]{2006ApJS..167...59H,2010ApJ...709..680H}. Also, a high effective temperature of the SSS indicates a high-mass WD \citep{2005A&A...439.1061S}. A massive WD is required to produce recurrence times that are short enough to be observed \citep{1986ApJ...308..721T,2005ApJ...623..398Y}. The shortest time between two confirmed nova outbursts was eight years for the Galactic RN U~Sco \citep{2010ApJS..187..275S}. In the single-degenerate scenario of Type Ia supernova (SN~Ia) progenitors, RNe correspond to the final stage of the binary evolution \citep[see][]{1999ApJ...519..314H,1999ApJ...522..487H}.

Novae in our neighbour galaxy \m31 \citep[distance 780 kpc;][]{1998AJ....115.1916H,1998ApJ...503L.131S} have been observed for almost a century. Currently, almost a thousand nova outbursts are known in \m31 \citep[see the online catalogue\footnote{http://www.mpe.mpg.de/$\sim$m31novae/opt/m31/index.php} of][]{2007A&A...465..375P}. Dedicated X-ray monitoring observations of \m31 novae together with archival studies compiled a sample of 79 novae with SSS counterparts \citep[see][hereafter \mdk, and references therein]{2013arXiv1312.1241H}. A statistical analysis of this sample revealed strong correlations between optical and X-ray parameters, showing that novae that decline fast in the optical are hot in X-rays with short SSS durations (\mdk).

Nova \nova was first discovered in Dec 2008 by F. Kabashima \& K. Nishiyama\footnote{http://www.cbat.eps.harvard.edu/CBAT\_M31.html\#2008-12a}. Two additional eruptions were found in Oct 2011 and Oct 2012, with the latter being classified as the outburst of a He/N nova by \citet{2012ATel.4503....1S}. For a comprehensive description of the optical properties and the identification of the latter outbursts with \nova see the accompanying Letter by \citet[][hereafter \optk]{2014arXiv1401.2905D}. In Nov 2013, another outburst was detected by \citet{2013ATel.5607....1T} and this Letter focusses on the results from the subsequent X-ray monitoring. 

%
\section{Observations and data analysis}
\label{sec:obs}
%
Soon after the announcement of the Nov 2013 outburst of \nova we requested a target of opportunity (ToO) X-ray monitoring with \swift \citep{2004ApJ...611.1005G}. The extremely short recurrence time suggested a high-mass WD, the SSS phase of which we aimed successfully to detect and characterise. All observations are summarised in Table\,\ref{tab:obs}. Additionally, in Table\,\ref{tab:obs12} we list all observations of a \swift monitoring programme after the 2012 outburst that did not detect any X-ray source.

The \swift X-ray telescope \citep[XRT;][]{2005SSRv..120..165B} data analysis was based on the cleaned event files. We derived the count rates given in Table\,\ref{tab:obs} using the source statistics (\texttt{sosta}) tool within the HEAsoft XIMAGE package (version 4.5.1.). This method takes into account the background, estimated in a source-free region around the object, as well as corrections for sampling dead time, vignetting and point spread function (PSF). The upper limits in Tables\,\ref{tab:obs} and \ref{tab:obs12} were computed assuming a PSF constructed from the merged data of all detections.

The source and background photons were extracted from the event files within the HEAsoft Xselect environment (version v2.4c). The spectral analysis used XSPEC \citep[][version 12.8.1g]{1996ASPC..101...17A} with the T\"ubingen-Boulder ISM absorption model (\texttt{TBabs} in XSPEC), the photoelectric absorption cross-sections from \citet{1992ApJ...400..699B} and the ISM abundances from \citet{2000ApJ...542..914W}. All spectra were binned to include at least one photon per bin and fitted in XSPEC assuming Poisson statistics according to \citet{1979ApJ...228..939C}.

The ultraviolet (UV) magnitudes are given in the Vega system and were determined using the \texttt{uvotsource} tool. All magnitudes assume the \swift UV/optical telescope \citep[UVOT,][]{2005SSRv..120...95R} photometric system \citep{2008MNRAS.383..627P} and have not been corrected for extinction.

All upper limits correspond to $3\sigma$ confidence and all error ranges to $1\sigma$ confidence unless otherwise noted.

\section{Results}
\label{sec:results}
%
Nova \nova was clearly detected by \swift as a bright X-ray source following its 2013 outburst \citep[first announced by][]{2013ATel.5627....1H}. Based on the unprecedentedly short recurrence time we had expected a massive WD that should display a fast SSS appearance \citep[according to][\mdk]{2010ApJ...709..680H}. Consequently, the \swift monitoring was designed to start as soon as possible to constrain the beginning of the SSS phase. Surprisingly, the SSS turned on even faster than expected and the first observation, approximately six days after discovery, detected a bright SSS. The source was monitored with \swift until the SSS turned off two weeks later (see Table\,\ref{tab:obs}).

The source position, as determined from the merged XRT event files of all detections, is RA = 00h45m29.29s, Dec = +41$\degr$54$\arcmin 08\,\farcs5$ (J2000; 90\% confidence error $3\farcs6$). Although this is $4\farcs8$ away from the optical coordinates reported by \citet{2013ATel.5607....1T} (RA = 00h45m28.89s, Dec = +41$\degr$54$\arcmin 10\,\farcs2$) both positions are consistent ($\sim2\sigma$ level) due to the large uncertainties. Moreover, the appearance of the X-ray source soon after the optical nova outburst as well as its characteristics (see below) leave little doubt that both events are related. Therefore, we identify the transient X-ray source as the counterpart of \novak.

The best-fit parameters for a simultaneous fit of all X-ray spectra with a black body model are $kT = (97^{+5}_{-4})$ eV and \nh = ($1.4^{+0.2}_{-0.3}$) \hcm{21}. A $\chi^2$ test statistic for this fit gave a reduced $\chi^2$ of $1.19$ for 574 degrees of freedom. The estimated $kT$ is exceptionally high for an \m31 nova (cf. \mdk). From the \m31 reddening maps of \citet{2009A&A...507..283M} we derived an approximate E(B-V) $\sim0.26$ for the position of \novak. This corresponds to \nh $\sim 1.8$ \hcm{21} \citep[using the relation of][]{2009MNRAS.400.2050G}, which is consistent with the best-fit value.

\begin{figure}[t!]
  \resizebox{\hsize}{!}{\includegraphics[angle=0]{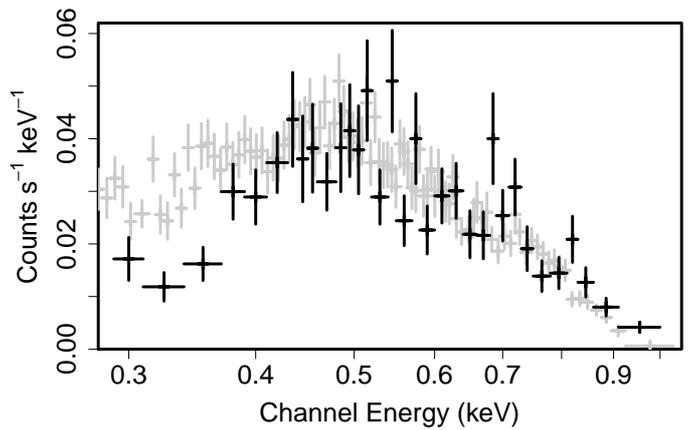}}
  \caption{\swift XRT spectrum extracted from merged event files. This is shown only for visualising the high effective temperature. Model fitting was performed on the individual spectra. In grey we show for comparison a scaled spectrum of the bright and hot nova M31N~2007-12b \citep{2011A&A...531A..22P} which is a suspected RN \citep{2009ApJ...705.1056B}.}
  \label{fig:xray_spec}
\end{figure}

A spectrum extracted from the combined event files of all detections is shown in Fig.\,\ref{fig:xray_spec}. This spectrum was grouped to include at least 20 counts per bin and is only used for visualisation. Practically all source photons have energies below 1~keV. This source is clearly a SSS. Because the Wien tail of the spectrum extends to (relatively) high energies we can infer that the SSS is unusually hot even without the use of specific spectral models \citep[which are controversial, see e.g.][]{2013A&A...559A..50N}.

\begin{figure}[t!]
  \resizebox{\hsize}{!}{\includegraphics[angle=0]{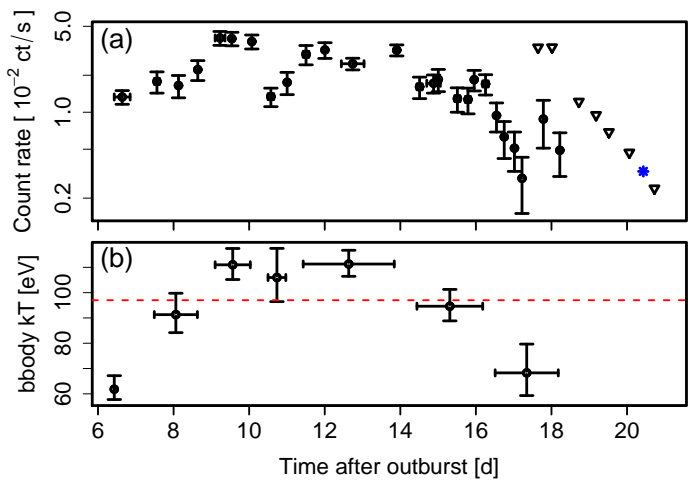}}
  \caption{Evolution of X-ray count rate (a) and effective black body temperature (b) of \nova during the 2013 \swift monitoring. This assumes an outburst on 2013-11-26.60 UT (see also Table\,\ref{tab:obs}). The error bars in time represent the duration of the individual observation (a) or the time between the grouped observations (b). \textit{Panel a:} Upper limits are indicated by black open triangles. A blue asterisk marks the first upper limit estimated from the \swift monitoring after the 2012 outburst (see Table\,\ref{tab:obs12}). \textit{Panel b:} Groups of observations have been fitted simultaneously with the \nh fixed to the global best fit. The red dashed line shows the overall best-fit temperature derived in Sect.\,\ref{sec:results}.}
  \label{fig:lc}
\end{figure}

In Fig.\,\ref{fig:lc}a we show the X-ray light curve of the nova during the 2013 outburst. This plot and Table\,\ref{tab:obs} assume that the optical outburst occurred between the last upper limit and the first detection reported by \citet{2013ATel.5607....1T}, i.e on Nov 26.6 2013 (with an uncertainty of $\pm0.5$~d). Following the first detection of the SSS on day six after the outburst its count rate increased and subsequently showed significant variability (see Fig.\,\ref{fig:lc}a). This variation was not due to instrumental vignetting or bad columns. The count rate started dropping on day 16 and the source disappeared a few days later. The last detection in an individual observation on day 18 showed a count rate that was approximately an order of magnitude below the peak level (see also Table\,\ref{tab:obs}).

The SSS might still have been detected at the $2\sigma$ level in the merged 14.3~ks of the last five upper limit observations. However, with ($7.7\pm3.4$) \tpower{-4} ct s$^{-1}$ its count rate would have been more than an order of magnitude below the bright phase. Therefore, this bright SSS phase which indicates stable hydrogen burning \citep[see e.g.][]{2005A&A...439.1061S}, was definitely over at this point. Consequently, we assume a SSS turn-off time (\toffk) of $(19\pm1)$~d. This is consistent with the results of the low-cadence \swift monitoring following the 2012 eruption that failed to detect a SSS $\sim20$~d after outburst (see Table\,\ref{tab:obs12} and Fig.\,\ref{fig:lc}a).

The X-ray variability might have been accompanied by spectral variation. In Fig.\,\ref{fig:lc}b we show the results of a simultaneous fitting (with \nh fixed to the global best fit) of separate groups of consecutive observations with similar count rate. Figure\,\ref{fig:lc}b shows indications for lower effective temperatures at the very beginning and end of the SSS phase. We also fitted the second and sixth temperature bin (low count rates) and the third and fifth bin (high count rates) in Fig.\,\ref{fig:lc}b simultaneously with fixed absorption. These groups of observations represent the high- and low-luminosity states of \novak, with the forth temperature bin possibly belonging to a transitory stage. The resulting best-fit temperatures were found to be different, albeit with weak significance: $kT$ = $(105\pm3)$ eV (high) vs $(91\pm4)$ eV (low). 

Furthermore, we found significant X-ray variability within the first (6~ks) \swift observation of the 2013 outburst \citep[see also][]{2013ATel.5633....1H}. This count rate variation was accompanied by variability in the UV. Subsequently, no significant UV variations were found against a gradual decline (see Table\,\ref{tab:obs}). In Fig.\,\ref{fig:xray_uv} we show both short-term light curves based on the individual \swift snapshots during the first detection. They appear to be anti-correlated. This process could indicate the gradual emergence of the SSS. Therefore, we estimate a SSS turn-on time (\tonk) of $(6\pm1)$~d.

\section{Discussion}
\label{sec:discuss}
%
\subsection{Variability}
\label{sec:disc_var}

\begin{figure}[t!]
  \resizebox{\hsize}{!}{\includegraphics[angle=0]{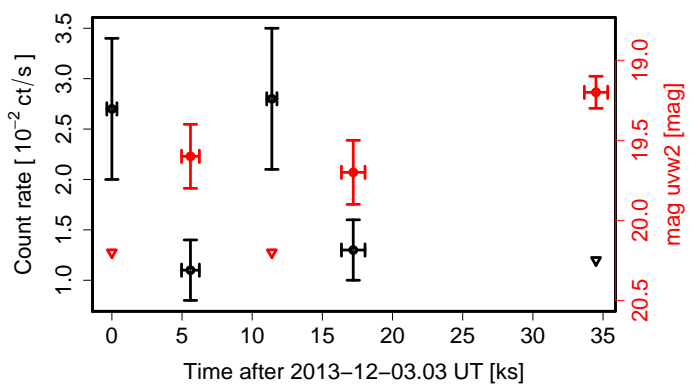}}
  \caption{Short-term X-ray and UV light curves for the five snapshots of \swift observation 00032613005 on day six after outburst (see Table\,\ref{tab:obs}).}
  \label{fig:xray_uv}
\end{figure}

Significant X-ray variability during the early stages of the SSS phase has been observed in several Galactic novae \citep[e.g. KT~Eri, RS~Oph;][]{2010ATel.2392....1B,2011ApJ...727..124O}. One interpretation of this effect is a variable absorption column during the emergence of the SSS from the expanding ejecta and/or optically thick wind \citep[see e.g.][]{2006ApJS..167...59H}. Short-term X-ray variability of various kinds has been observed in the \m31 nova sample (see \mdk).

The anti-correlation between X-ray and UV flux during the early SSS phase (Fig.\ref{fig:xray_uv}) and the later possible temperature variations (Fig.\,\ref{fig:lc}b) might point towards intrinsic variations in the WD temperature. \citet{2011ApJS..197...31S} suggested a similar scenario to explain the anti-correlated X-ray and UV variability during the SSS phase of the Galactic nova V458~Vul \citep[see also][]{2009AJ....137.4160N}. They drew an interesting parallel to the variability in persistent SSSs which might originate in accretion-rate variations that cause the WD photosphere to expand or shrink \citep[see][]{2000A&A...354L..37R,2002A&A...387..944G}. Based on its frequent outbursts and evidence from the optical data (\optk) \nova is likely to have a high accretion rate. The observed variability might indicate that accretion had been re-established as early as six days after outburst.

\subsection{Theoretical implications of the short recurrence time}
\label{sec:disc_theo}
Theoretically, a recurrence period ($\tau_r$) of order one year is only expected for extremely massive WDs ($\gtrsim 1.35$\msun) with a high mass accretion rate exceeding $10^{-7}$ \msun~yr$^{-1}$ \citep[][Saio et al. 2014, in prep.]{2013ApJ...777..136W}. A massive WD is consistent with the high temperature suggested in our X-ray spectra (Sect.\,\ref{sec:results}, Fig.\,\ref{fig:xray_spec}). For such a short $\tau_r$, however, the accreted mass is too small (several $10^{-7}$ \msun) for the pseudo-photosphere to bloat to a red-giant size. Therefore, we would expect a relatively high effective surface temperature even at the optical maximum that results in a faint peak magnitude. The shorter the $\tau_r$, the fainter the peak magnitude. 

The low peak brightness of \nova (never above 18th mag, see \optk) seems consistent with this scenario. While high accretion rates can also trigger hydrogen shell flashes on less massive WDs without causing a nova \citep[see][]{1993A&A...269..291J}, the observed evidence suggests that \nova was experiencing actual nova outbursts. Optically faint novae with fast and hot SSS phase might generally be good RN candidates.

An extremely massive WD can be born as an ONe WD in the final stage of stellar evolution for a narrow range of zero-age main-sequence mass ($\sim 8$--$10$~\msun). Another possibility is mass-accretion onto an initially less massive WD in a binary system. If a (CO) WD was born with $M_{\rm WD} < 1.07$~\msun\, and accumulates mass until reaching the Chandrasekhar mass it will explode as a SN~Ia. RNe with short $\tau_r$ correspond to immediate progenitors of SNe~Ia in the single degenerate scenario \citep{1999ApJ...519..314H,1999ApJ...522..487H}.

\subsection{Population picture}
\label{sec:disc_pop}
\begin{figure}[t!]
  \resizebox{\hsize}{!}{\includegraphics[angle=0]{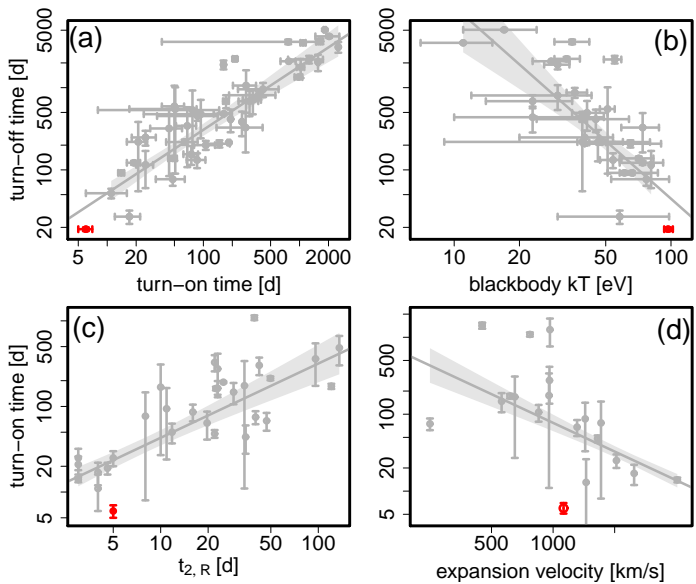}}
  \caption{Double-logarithmic plots of the nova parameter correlations from \mdk. The solid grey lines indicate a robust powerlaw fit with corresponding 95\% confidence regions in light grey. The correlations displayed are: (a) \ton vs \toffk, (b) black body $kT$ vs \toffk, (c) $R$ band optical decay time $t_{2,R}$ vs \tonk, and (d) expansion velocity vs \tonk. All time scales are in units of days after outburst. Overplotted in red are the parameters of \nova as derived in Sect.\,\ref{sec:results} or by \optk.}
  \label{fig:corr}
\end{figure}

In Fig.\,\ref{fig:corr} we show the position of \nova in the five-parameter space of the correlations presented by \md for the \m31 nova sample. The SSS time scales were unprecedentedly short, also compared to Galactic novae where e.g. the fastest \ton was displayed by the RN U~Sco \citep[$\gtrsim 8$~d;][]{2010ATel.2430....1S}. The average SSS temperature seems considerably higher than measured for any other \m31 nova and might have been even hotter during the high-count rate observations (see Sect.\,\ref{sec:results} and Fig.\,\ref{fig:lc}b). Overall, the X-ray parameter relations (Figs.\,\ref{fig:corr}a and b) are broadly consistent with the population trends. \nova therefore provides an important extension of these relations into previously unpopulated regions of the parameter space.

The two optical parameters, \ttwo and \vexp (see \optk), show stronger deviations from the rest of the nova sample. Such a behaviour might be consistent with a fainter nova outburst as suggested by theoretical models (Sect.\,\ref{sec:disc_theo}). However, for the two correlations in Figs.\,\ref{fig:corr}c and d the scatter is generally larger. At this stage, it is difficult to determine whether the connection between the optical and X-ray parameters of \nova is significantly different from the overall population trend.

%

\subsection{Previous X-ray outbursts}
\label{sec:disc_prev}
During a literature search we found that \nova had actually been detected in X-rays before the first optical outburst was reported. It was initially discovered as the ``recurrent supersoft X-ray transient'' RX~J0045.4+4154 by \citet{1995ApJ...445L.125W} based on archival ROSAT data. Because no optical nova was known before 2008, this source was not recognised as a nova counterpart by \citet{2005A&A...442..879P} in the course of their archival analysis. \citet{1995ApJ...445L.125W} reported two X-ray outbursts in Feb 1992 and Jan 1993. The recurrence time of approximately one year as well as the short duration of the outbursts and the high black body temperature ($\sim90$~eV) estimated by \citet{1995ApJ...445L.125W} are well in agreement with the results reported here. 

Furthermore, another outburst of this ``known recurrent transient'' was reported by \citet{2004ApJ...609..735W} in \chandra data taken in Sep 2001. Because this observation used the HRC detector no X-ray spectrum exists. The outburst was shorter than five months. Overall, there is strong evidence that \nova had at least three outbursts before 2008. A follow-up paper will study the optical and X-ray archives in more detail.

\section{Conclusions}
\label{sec:summary}
%
The X-ray properties of \nova suggest coherently that the binary system hosts a high-mass WD. The fast recurrence times and optical evidence indicate a high accretion rate the re-establishing of which could explain the observed early X-ray and UV variability. \m31 remains a treasure island for nova science and might hold the key to the understanding of the (potentially numerous) optically faint RNe and their impact on SN~Ia progenitor theories. The position of \nova will be watched closely (in particular) in late 2014.

\begin{acknowledgements}
We thank the referee, Eric Schlegel, for a careful reading of the manuscript. We wish to thank Margarita Hernanz for constructive comments on an early version of this Letter. We are grateful to the \swift Team for making the ToO observations possible, in particular N. Gehrels, the duty scientists as well as the science planners. M.H. acknowledges support from an ESA fellowship. S.C.W. acknowledges PhD funding from STFC. A.W.S. acknowledges support from NSF grant AST1009566.
\end{acknowledgements}

\bibliographystyle{aa}

%
\onltab{
\begin{table*}[ht]
\caption{\swift observations of nova \novak.}
\label{tab:obs}
\begin{center}
\begin{tabular}{rrrrrrrrrrrr}\hline\hline \noalign{\smallskip}
ObsID & Exp$^a$ & Date$^b$ & MJD$^b$ & $\Delta t^c$ & \multicolumn{3}{c}{UV$^d$ [mag]} & Rate &  L$_{0.2-1.0}$ $^e$\\
& [ks] & [UT] & [d] & [d] & uvm2 & uvw1 & uvw2 & [\power{-2} ct s$^{-1}$] & \power{38} erg s$^{-1}$]\\ \hline \noalign{\smallskip}
00032613005 & 6.0 & 2013-12-03.03 & 56629.03 & 6.43 & - & - & $19.4\pm0.1$ & $1.3\pm0.2$ & $1.0\pm0.1$ \\
00032613006 & 2.0 & 2013-12-04.09 & 56630.09 & 7.49 & $19.8\pm0.3$ & $>19.5$ & $19.7\pm0.2$ & $1.8\pm0.4$ & $1.3\pm0.3$ \\
00032613007 & 2.0 & 2013-12-04.70 & 56630.70 & 8.10 & $19.4\pm0.2$ & $19.4\pm0.2$ & $19.3\pm0.2$ & $1.7\pm0.3$ & $1.2\pm0.3$ \\
00032613008 & 1.7 & 2013-12-05.23 & 56631.23 & 8.63 & $20.0\pm0.4$ & $19.6\pm0.2$ & $19.8\pm0.3$ & $2.2\pm0.4$ & $1.7\pm0.3$ \\
00032613009 & 1.9 & 2013-12-05.70 & 56631.70 & 9.10 & $19.8\pm0.3$ & $>19.9$ & $19.9\pm0.3$ & $4.0\pm0.5$ & $3.0\pm0.4$ \\
00032613010 & 2.0 & 2013-12-06.10 & 56632.10 & 9.50 & $19.9\pm0.3$ & $19.8\pm0.3$ & $19.5\pm0.2$ & $4.0\pm0.5$ & $3.0\pm0.4$ \\
00032613011 & 2.0 & 2013-12-06.63 & 56632.63 & 10.03 & $>20.2$ & $19.7\pm0.3$ & $20.0\pm0.3$ & $3.8\pm0.5$ & $2.8\pm0.4$ \\
00032613012 & 3.3 & 2013-12-07.10 & 56633.10 & 10.50 & $20.1\pm0.2$ & $>19.8$ & $20.2\pm0.3$ & $1.3\pm0.2$ & $1.0\pm0.2$ \\
00032613013 & 1.8 & 2013-12-07.57 & 56633.57 & 10.97 & $20.1\pm0.4$ & $19.4\pm0.2$ & $19.6\pm0.2$ & $1.7\pm0.4$ & $1.3\pm0.3$ \\
00032613014 & 1.5 & 2013-12-08.03 & 56634.03 & 11.43 & $19.6\pm0.3$ & $>20.0$ & $19.9\pm0.3$ & $3.0\pm0.5$ & $2.2\pm0.4$ \\
00032613015 & 1.8 & 2013-12-08.57 & 56634.57 & 11.97 & $19.9\pm0.4$ & $20.0\pm0.3$ & $>20.3$ & $3.2\pm0.5$ & $2.4\pm0.4$ \\
00032613016 & 4.4 & 2013-12-09.03 & 56635.04 & 12.44 & $19.8\pm0.2$ & $19.7\pm0.3$ & $20.0\pm0.2$ & $2.5\pm0.3$ & $1.9\pm0.2$ \\
00032613017 & 3.8 & 2013-12-10.44 & 56636.44 & 13.84 & $>20.6$ & $19.9\pm0.2$ & $20.0\pm0.2$ & $3.2\pm0.3$ & $2.4\pm0.2$ \\
00032613018 & 2.1 & 2013-12-11.04 & 56637.04 & 14.44 & $20.0\pm0.4$ & $>20.2$ & $>20.4$ & $1.6\pm0.3$ & $1.2\pm0.2$ \\
00032613019 & 3.0 & 2013-12-11.31 & 56637.30 & 14.70 & $20.1\pm0.3$ & $19.6\pm0.3$ & $20.2\pm0.3$ & $1.7\pm0.3$ & $1.3\pm0.2$ \\
00032613020 & 1.9 & 2013-12-11.57 & 56637.57 & 14.97 & $>20.0$ & $19.6\pm0.3$ & $19.9\pm0.3$ & $1.9\pm0.4$ & $1.4\pm0.3$ \\
00032613021 & 1.9 & 2013-12-12.04 & 56638.04 & 15.44 & $19.9\pm0.2$ & - & - & $1.3\pm0.3$ & $1.0\pm0.2$ \\
00032613022 & 1.9 & 2013-12-12.32 & 56638.32 & 15.72 & $20.4\pm0.3$ & - & - & $1.3\pm0.3$ & $1.0\pm0.2$ \\
00032613023 & 1.9 & 2013-12-12.51 & 56638.52 & 15.92 & $20.2\pm0.4$ & $19.8\pm0.3$ & $20.0\pm0.3$ & $1.8\pm0.4$ & $1.4\pm0.3$ \\
00032613024 & 2.1 & 2013-12-12.78 & 56638.78 & 16.18 & $>20.9$ & - & - & $1.7\pm0.3$ & $1.3\pm0.2$ \\
00032613025 & 2.0 & 2013-12-13.10 & 56639.11 & 16.51 & - & $19.8\pm0.2$ & - & $0.9\pm0.2$ & $0.7\pm0.2$ \\
00032613026 & 1.9 & 2013-12-13.31 & 56639.31 & 16.71 & - & $20.4\pm0.3$ & - & $0.6\pm0.2$ & $0.5\pm0.2$ \\
00032613027 & 1.9 & 2013-12-13.58 & 56639.58 & 16.98 & - & $20.2\pm0.2$ & - & $0.5\pm0.2$ & $0.4\pm0.1$ \\
00032613028 & 2.0 & 2013-12-13.78 & 56639.78 & 17.18 & $>20.2$ & $>20.2$ & $>20.5$ & $0.3\pm0.1$ & $0.2\pm0.1$ \\
00032613029 & 0.3 & 2013-12-14.24 & 56640.24 & 17.64 & - & - & - & $<3.4$ & $<2.5$ \\
00032613030 & 0.9 & 2013-12-14.38 & 56640.38 & 17.78 & - & - & - & $0.9\pm0.4$ & $0.7\pm0.3$ \\
00032613031 & 1.9 & 2013-12-14.52 & 56640.52 & 17.92 & - & - & - & $<3.4$ & $<2.5$ \\
00032613032 & 1.9 & 2013-12-14.78 & 56640.78 & 18.18 & $>20.2$ & $20.1\pm0.4$ & $20.3\pm0.3$ & $0.5\pm0.2$ & $0.4\pm0.1$ \\
00032613036 & 1.6 & 2013-12-15.06 & 56641.06 & 18.46 & $>20.1$ & $>20.0$ & $>20.3$ & $<1.2$ & $<0.9$ \\
00032613033 & 1.4 & 2013-12-15.64 & 56641.64 & 19.04 & - & - & $20.8\pm0.3$ & $<0.9$ & $<0.7$ \\
00032613034 & 1.9 & 2013-12-16.05 & 56642.05 & 19.45 & $>20.2$ & $>20.1$ & $>20.4$ & $<0.7$ & $<0.5$ \\
00032613035 & 1.7 & 2013-12-16.52 & 56642.52 & 19.92 & $20.7\pm0.3$ & - & - & $<0.5$ & $<0.3$ \\
00032613037 & 7.6 & 2013-12-17.06 & 56643.06 & 20.46 & $>19.9$ & $20.4\pm0.2$ & $>20.3$ & $<0.2$ & $<0.2$ \\
00032613038 & 7.8 & 2013-12-20.66 & 56646.66 & 24.06 & $20.4\pm0.2$ & $>20.6$ & $20.2\pm0.3$ & $<0.1$ & $<0.1$ \\
00032613039 & 8.0 & 2013-12-23.26 & 56649.26 & 26.66 & $20.0\pm0.4$ & $>19.6$ & $20.5\pm0.2$ & $<0.2$ & $<0.1$ \\
00032613040 & 3.8 & 2013-12-23.59 & 56649.59 & 26.99 & - & - & - & $<0.2$ & $<0.2$ \\
\hline
\end{tabular}
\end{center}
\noindent
Notes:\hspace{0.1cm} $^a $: Dead-time corrected exposure time; $^b $: Start date of the observation; $^c $: Time in days after the outburst of nova \nova in the optical on 2013-11-26.60 UT \citep[MJD = 56622.60; see][]{2013ATel.5607....1T}; $^d $: \swift UVOT filters were UVM2 (166-268nm), UVW1 (181-321nm) and UVW2 (112-264nm); $^e $:X-ray luminosities (unabsorbed, blackbody fit, 0.2 - 1.0 keV) and upper limits were estimated according to Sect.\,\ref{sec:results}.\\
\end{table*}
}
%

%
\onltab{
\begin{table*}[ht]
\caption{\swift observations of the 2012 outburst of nova \novak.}
\label{tab:obs12}
\begin{center}
\begin{tabular}{rrrrrrrrrrr}\hline\hline \noalign{\smallskip}
ObsID & Exp$^a$ & Date$^b$ & MJD$^b$ & $\Delta t^c$ & \multicolumn{3}{c}{UV$^d$ [mag]} & Rate &  L$_{0.2-1.0}$ $^e$\\
& [ks] & [UT] & [d] & [d] & U & uvw1 & uvw2 & [\power{-2} ct s$^{-1}$] & \power{38} erg s$^{-1}$]\\ \hline \noalign{\smallskip}
00032613001 & 4.0 & 2012-11-06.45 & 56237.45 & 20.35 & - & - & $20.2\pm0.2$ & $<0.3$ & $<0.2$ \\
00032613002 & 4.2 & 2012-11-16.34 & 56247.34 & 30.24 & - & $19.8\pm0.1$ & - & $<0.3$ & $<0.2$ \\
00032613003 & 3.5 & 2012-12-16.13 & 56277.13 & 60.03 & - & - & $20.4\pm0.2$ & $<0.3$ & $<0.2$ \\
00032613004 & 3.7 & 2013-02-05.36 & 56328.36 & 111.26 & $19.8\pm0.2$ & - & - & $<0.4$ & $<0.3$ \\
\hline
\end{tabular}
\end{center}
\noindent
Notes:\hspace{0.1cm} As for Table\,\ref{tab:obs}. The time $^c $ is in days after the outburst of nova \nova in the optical on 2012-10-17.10 UT (MJD = 56217.10; see Sect.\,\ref{sec:results}.\\
\end{table*}
}

\end{document}